\documentclass[%
reprint,
superscriptaddress,
 amsmath,amssymb,
 aps,prl
]{revtex4-1}

\usepackage{graphicx}% Include figure files
\usepackage{dcolumn}% Align table columns on decimal point
\usepackage{bm}% bold math
\usepackage{hyperref}% add hypertext capabilities
\usepackage[usenames,dvipsnames]{color}
\usepackage{listings}
\begin{document}

%\preprint{APS/123-QED}

\title{Object-oriented Pseudo-spectral code TARANG for turbulence simulation}
\author{Mahendra K. Verma}
\affiliation{Department of Physics, Indian Institute of Technology, Kanpur, India 208016}
 \email{mkv@iitk.ac.in}

\date{\today}% It is always \today, today,

\begin{abstract}
In this paper we describe the design and implementation of TARANG, a pseudospectral code to simulate turbulent flows in fluids, magnetohydrodynamics (MHD), convection, passive scalar, etc.   We use the object-oriented features of C++ to abstract operations involved in the simulation.  TARANG has been validated and used for solving problems in convection and MHD.
\end{abstract}

\maketitle
%\tableofcontents

\section{\label{sec:Introduction}  Introduction}

Turbulence is one of the most challenging and unsolved problems of physics.  Theoretical or exact analysis of turbulence has been rare due to the complex nonlinearities present in the equations.  Therefore, major attempts to understand turbulence has been through experiments and numerical simulations.  Over time very powerful supercomputers and software tools have emerged, that have propelled the computational capabilities of turbulent flows to unimaginable heights.  

Some of the popular schemes to solve fluid flows are {\em finite difference, finite element, finite volume,  spectral elements, pseudo-spectral, vortex method}, etc.  The pseudo-spectral method~\cite{Canuto:book1} is most accurate among them, and it employed for studying small-scale turbulence.  In the present paper we will describe the design and capability of a pseudo-spectral code named ``TARANG'', which has been developed by our group.   TARANG is a Sanskrit word that means ``waves".

TARANG is an object-oriented parallel pseudo-spectral code that can simulate flows in fluids, magnetofluids, convection, magnetoconvection etc.  The convective flow module can also be used to solve Rayleigh-Taylor instability and turbulence, stratified flows, nonBoussinesq convection, and passive scalars, etc.   We designed TARANG in an object-oriented fashion for generality and easy adoptability to solve varied problems. We  make use of fast libraries, FFTW (Fastest Fourier Transform in the West) and blitz++ for efficiency.  The code is neatly segregated into libraries and src directory.  The fluid, magnetohydrodymics (MHD), convection solvers etc. make use of the libraries, and they are arranged in the src directory.

We will describe the design issues and various features of TARANG in the following sections.

\section{\label{sec:design}  Design and implementation issues of TARANG}
In TARANG we use the object-oriented features of C++ to design general purpose libraries to create solvers for the incompressible fluid flows.  For example, a library function {\em Compute\_nlin} computes ${\bf (u \cdot \nabla) u}$ for all basis functions.  A solver containing many complex features and boundary conditions, and more field variables are easily constructed using these functions.   Main features including the class structures of TARANG are described below.

\subsection{\label{sec:ext_lib}External libraries}
The handling of arrays and mathematical functions is computationally slow in C++.  Fortunately, several efficient C++ libraries are available to perform the above tasks.  We chose {\em blitz}++ for TARANG since it handles multidimensional arrays in a nice and succinct manner.  The other libraries with similar functions are {\em boost, ndarrays}, and {\em eigen}, but presently we continue our development with blitz++.

Pseudospectral codes  use FFT (Fast Fourier Transforms) heavily.  In a typical code, approximately 80\% of the computational time is spent of FFT.   FFTW is the most popular and the most efficient parallel FFT library available today, therefore we use FFTW in our code.  

\subsection{\label{sec:basis} Basis functions}
A pseudo-spectral code uses basis functions to expand the real-space functions.  The choice of the basis functions depend critically on the boundary conditions.   $\exp{(i {\bf k \cdot x})}$, where ${\bf k,x}$ are the wavenumber and real-space coordinates respectively, is the natural choice for periodic boundary conditions.  Convection, channel flows etc.~however involves walls. All the components of the velocity fields at the wall must vanish for the no-slip boundary condition.  Chebyshev and Legandre polynomials are used to expand such functions.  For the free-slip boundary condition, the velocity field perpendicular to the wall, and the perpendicular gradient of the horizontal velocity components are zero.  Sine and cosine functions are obvious and simple choices for such simulations.  Spherical harmonics are natural choice for spherical simulations.

TARANG focusses on basis-independent libraries, so we have designed the basis functions in a modular fashion.
At present TARANG has FOUR and SCFT (sin/cos-Fourier) basis functions that can simulate flows in a box geometry under periodic boundary conditions (all directions) and free-slip boundary conditions along $x$, and periodic along $y$ and $z$ directions.    The SCFT basis functions with appropriate modifications has been also used to simulate flows with free-slip boundary conditions along all the directions.  The no-slip boundary conditions has been successfully tested for channel flow, but this module will be integrated with TARANG soon.  The spherical geometry and cylindrical geometry will be implemented in future.

Primary functions related to the basis functions are {\em forward\_transform} (transformation from the real space to Fourier space), {\em inverse\_transform} (transformation from the Fourier space to the real space), computation of energy spectrum etc.  We use FFTW for majority of transform operations.

\subsection{\label{Objects} Objects of TARANG}

\begin{figure}[b]
\includegraphics[scale=0.75]{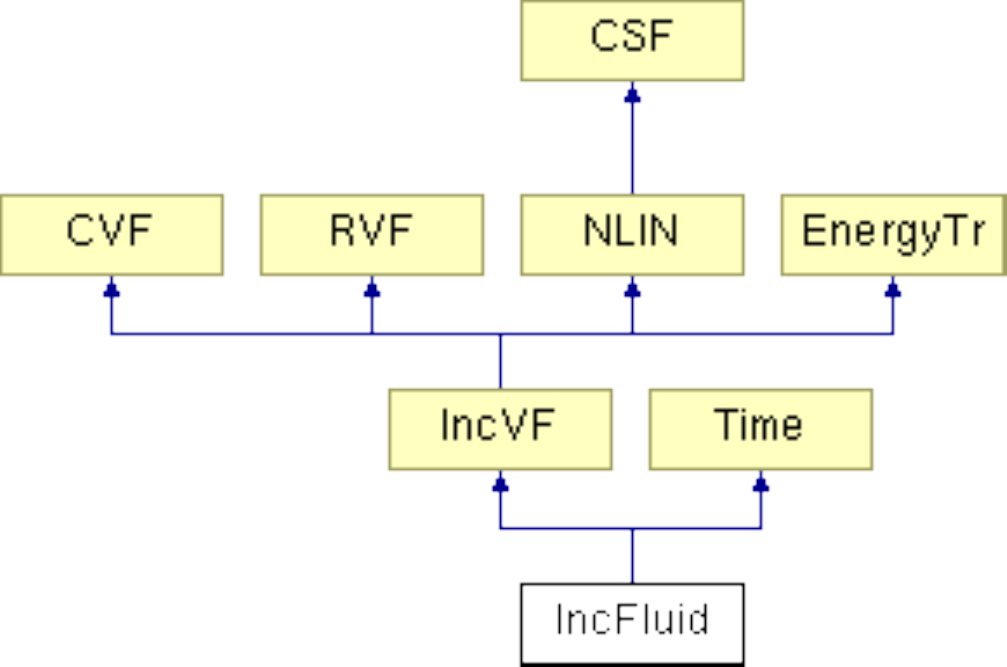}% Here is how to import EPS art
\caption{The class structure of IncFluid (Incompressible Fluids), which is the final  class of TARANG.}
\label{fig:IncFluid}
\end{figure}

The class structure of TARANG is illustrated in Fig.~\ref{fig:IncFluid}.  The class {\em IncFluid} contains the Incompressible velocity field and the associate solvers.  This is the final class that inherits more classes.  We describe the main features of the classes below.

\begin{figure}[b]
\includegraphics{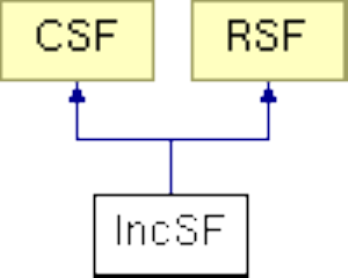}% Here is how to import EPS art
\caption{ The class structure of IncSF (Incompressible scalar field).}
\label{fig:IncSF}
\end{figure}

\subsubsection{Class CVF:  Complex Vector Field}
The class {\em CVF} stands for {\em Complex Vector Field}.  It contains three dynamic arrays associated with the three components of the velocity or magnetic fields in the Fourier space.  As mandated by FFTW, the size of each array is $N_1, N_2, N_3/2+1$ that spans wavenumbers $(k_1, k_2,k_3)  = [-N_1/2:N_1/2,-N_2/2:N_2/2, 0:N_3/2]$ in FOUR basis, and  $(k_1, k_2,k_3)  = [0:N_1,-N_2/2:N_2/2, 0:N_3/2]$  in SCFT basis.  These arrays contains complex numbers that represent the Fourier amplitude of the vector field.  The arrays are created dynamically at the run-time.

Forward and inverse transforms, and input/output of the vector fields are some of the main functions of the class {\em CVF}.

\subsubsection{Class RVF: Real Vector Field}
The class {\em RVF} stands for {\em Real Vector Field}, and it contains three dynamic arrays to represent the vector fields in the real space.  We still create complex arrays of the size $N_1, N_2, N_3/2+1$  for ease of FFTW and blitz++ operations; here the real and imaginary parts of a complex number represent two adjacent points in the real space.

\subsubsection{Class CSF: Complex Scalar Field}
The class {\em CSF} stands for {\em Complex Scalar Field}.  It contains a dynamic arrays associated with a scalar field, e.g., temperature in the Fourier space.  The indexing of the array is similar to that of {\em CVF}.

\subsubsection{Class RSF: Real Scalar Field}
The class {\em RSF} stands for {\em Real Scalar Field}, and it contains a dynamic array associate with a scalar.  The array features are same as that for {\bf RVF}.

The above four classes reside in directory named {\em fields}.  

\subsubsection{Class IncVF: Incompressible Vector Field}
The class {\em IncVF}, acronym for {\em Incompressible Vector Field}, contains most crucial functions of the solver. 
 {\em IncVF} inherits classes {\em CVF, RVF, NLIN}, {\em EnergyTr}.  The classes  {\em CVF, RVF} contain the velocity field in the Fourier and real space respectively.   The class {\em NLIN} contains three arrays for storing the nonlinear term $\widehat{\partial_j u_j u_i}$, where the symbol $\widehat{}$ represents the Fourier transform.  In addition, {\em NLIN} inherits {\em CSF}, whose array is used for storing the pressure field.  The class  {\em EnergyTr} contains function for computing energy flux, shell-to-shell energy transfer etc.~\cite{Dar:PD2001,Verma:PR2004}.
  
 The class {\em IncVF} also contains three arrays {\em Force\_i}  to store the force fields, and two array {\em *VF\_temp, *VF\_temp2} to save temporary fields.  It also has an array {\em *VF\_temp\_r} that is used for storing temporary arrays in real space.
 
 \subsubsection{Class IncSF: Incompressible Scalar Field}
 The class {\em IncSF} is used for a scalar field accompanying incompressible velocity field.  For example, in Rayleigh B\'{e}nard convection this class is used to represent the temperature field.   {\em IncSF} inherits a {\em CSF} and a {\em RSF} to store the scalar field in the Fourier and real space respectively (see FIg.~\ref{fig:IncSF}).  It also contains arrays {\em nlin, Force}, and {\em SF\_temp} to store nonlinear term ${\bf u \cdot \nabla} T$, forcing, and temporary array.
 
 \subsubsection{Compute\_nlin(): a class function}
 The class {\em IncVF} has many functions.  However, {\em Compute\_nlin} is one of the most important functions of this class.  We will describe this function as an illustration of TARANG function:

\lstset{caption=Compute\_nlin,label=compute_nlin}
\begin{lstlisting}
void IncVF::Compute_nlin() 
{					
	*V1r = *V1;  
	*V2r = *V2; 
	*V3r = *V3;	
	// Inverse transform of *Vir using 
	// *VF_temp_r as temporary array
	RV_Inverse_transform(*VF_temp_r);						
	// Vr[i] -> Vr[i]^2 stored in nlin[i]				                      
	Compute_RSprod_diag();										
	// nlin[i]= Di T[Vr[i]^2]; 
	// T = Forward transform
	// Di=derivative along i-th dirn
	NLIN_diag_Forward_transform_derivative(*VF_temp_r);			
	// Vr[i] = Vr[i]*Vr[j]
	Compute_RSprod_offdiag();	
	// Vr[i] = T(Vr[i]*Vr[j])																
	RV_Forward_transform_RSprod(*VF_temp_r);					
	// nlin[i] = Dj[T(Uj * Ui)]
	Derivative_RSprod_VV();										
}
\end{lstlisting}  
%
%\lstinputlisting{source_filename.py}
%
%\begin{verbatim}
%void IncVF::Compute_nlin() 
%{					
%	*V1r = *V1;  
%	*V2r = *V2; 
%	*V3r = *V3;	
%
%	// Inverse transform of *Vir using 
%	// *VF_temp_r as temporary array
%	RV_Inverse_transform(*VF_temp_r);						
%	
%	// Vr[i] -> Vr[i]^2 stored in nlin[i]				                      
%	Compute_RSprod_diag();										
%	 
%	// nlin[i]= Di T[Vr[i]^2]; 
%	// T = Forward transform
%	// Di=derivative along i-th dirn
%	NLIN_diag_Forward_transform_derivative(*VF_temp_r);			
%
%	// Vr[i] = Vr[i]*Vr[j]
%	Compute_RSprod_offdiag();	
%	
%	// Vr[i] = T(Vr[i]*Vr[j])																
%	RV_Forward_transform_RSprod(*VF_temp_r);					
%
%	// nlin[i] = Dj[T(Uj * Ui)]
%	Derivative_RSprod_VV();										
%}
%\end{verbatim}

The comments above the C++ statements explain the logic of the functions.  Related functions compute the nonlinear term in the presence of scalar field and another vector field.  
\begin{itemize}
\item void Compute\_nlin\_scalar(IncSF\& T): nlin\_i = $\widehat{\partial_j u_j u_i}$ and 
T.nlin\_i = $\widehat{\partial_j u_j F}$, where T.F is the scalar field.
			
\item void Compute\_nlin\_RB(IncSF\& T): same as Compute\_nlin\_scalar(IncSF\& T).
				
\item void Compute\_nlin(IncVF\& W): nlin\_i = $\widehat{(\partial_j u_j u_i - \partial_j w_j w_i)}$ and 
W.nlin\_i =  $\widehat{(\partial_j u_j w_i - \partial_j w_j u_i)}$, where $w_i$ is the vector field associated with the {\em IncVF} class {\em W}.

\item void Compute\_nlin(IncVF\& W, IncSF\& T): nlin\_i = $\widehat{(\partial_j u_j u_i - \partial_j w_j w_i)}$, and 
W.nlin\_i =  $\widehat{(\partial_j u_j w_i - \partial_j w_j u_i)}$, and T.nlin = $\widehat{\partial_j u_j F}$, with the same interpretation as given above.
\end{itemize}

\subsubsection{Class IncFluid: Incompressible Fluid}
The class {\em IncFluid} inherits {\em IncVF} and {\em Time}.  Major functions of this class deal with time advancement of solver, forcing function, and input/output.  The files and their associated functions are defined within this class.  At present, the code includes Euler, Runge-Kutta second order (RK2), and Runge-Kutta fourth order (RK4) for the time advance function.  The forcing function is used to include the buoyancy term in Rayleigh-B\'{e}nard convection (RBC), Coriolis force in the rotating turbulence etc.  

For input/output, we have the option of reading/writing the data either the ASCII format or in High Density Format(HDF5) format.  The HDF5 part of the code is being integrated with the main code.  Also note that the classes {\em IncVF} and {\em IncFluid} have multiple inheritance.  

\subsection{\label{Objects} Solvers of TARANG}
We invoke the library functions discussed above to create solvers for fluid, magnetohydrodynamics, passive scalar, RBC flows etc.  We illustrate a code segment containing the time-loop of fluid solver for  an illustration.

\lstset{caption=Fluid Solver,label=fluid_solver}
\begin{lstlisting}
//	A code segment of the fluid solver 
// Read initial condition	
	U.Read_init_cond();	
	int  iter=0;  // iterations 
	U.Tnow = U.Tinit;
	do 
	{
		U.Compute_force();	
		U.Compute_nlin();
		U.Add_force();		
		U.Compute_pressure();  
		U.Tdt = U.Get_dt();
		U.Tnow = U.Tnow + U.Tdt;
		iter++; 
		U.Time_advance();
		// FIELD AT new time
		U.Output_all_inloop();
	}
	while (U.Tnow < U.Tfinal);
\end{lstlisting}

In the above code segment, {\em U} is an instantiation of the class {\em IncFluid} which contains the incompressible velocity field.  Most of the functions are obvious.  The function {\em U.Get\_dt()} computes $dt$ using CFL condition.

For the RBC, the above code segment is modified slightly.  We create an instantiation $T$ of the class {\em IncSF} to represent the temperature field.

\lstset{caption=RBC solver,label=rbc_solver}
\begin{lstlisting}
// A code segment of the RBC solver 
// Read initial condition	
	U.Read_init_cond(T);	
	int  iter=0;  // iterations 
	U.Tnow = U.Tinit;
	do 
	{
		U.Compute_force(T);	
		U.Compute_nlin(T);
		U.Add_force(T);		
		U.Compute_pressure();  
		U.Tdt = U.Get_dt(T);
		U.Tnow = U.Tnow + U.Tdt;
		iter++; 
		U.Time_advance(T);
		// FIELD AT new time
		U.Output_all_inloop(T);
	}
	while (U.Tnow < U.Tfinal);
\end{lstlisting}

\section{Sample simulation results and validation}
We have performed simulations on fluids, convective, and MHD flows using TARANG 
on grids from $64^3$ to $1024^3$~\cite{Pal:EPL2009,Mishra:PRE2010,Mishra:EPL2010,Yadav:EPL2010}.   The reader is referred to the published work for the
scientific details.  In the following discussion we detail some of the validations we performed before we launched large simulations.

\subsection{Simulations of RBC}
Rayleigh-B\'enard convection (RBC) is an idealized version of the thermal convection in fluid. In the set up, a layer of incompressible fluid is confined between two thermally conducting plates separated by a distance $d$. The bottom plate is heated and an adverse temperature  gradient $\beta$ is set across the fluid layer. The system is governed by the following equations:
\begin{eqnarray}
\partial_{t}{\bf u} + ({\bf u}\cdot\nabla){\bf u} &=& -\nabla{p}+PR\theta\hat{z}+P\nabla^2 {\bf u}, \label{eq:NS}\\
\partial_{t}\theta + ({\bf u}\cdot\nabla)\theta &=& u_3 + \nabla^2 \theta, \label{eq:temp} \\
\nabla \cdot {\bf u} & = & 0, \label{eq:continuity} 
\end{eqnarray}
where ${\bf u}=(u_1,u_2,u_3)$  is the velocity field, $\theta$ is the
perturbation in the temperature field from the steady conduction
profile, and $\hat{z}$ is the vertically directed unit vector. The
equations are nondimensionalized by choosing length scale as $d$,
velocity scale  as $ d^{2}/\kappa$, and temperature
scale as $\beta d$, where  $\kappa$ is the thermal diffusivity of the fluid.  Two non
dimensional parameters in the equations are the Rayleigh number, $R=\alpha g \beta d^4/\nu \kappa$ and the Prandtl number, $P=\nu/\kappa$, where $\alpha$ is the coefficient of the volume
expansion, $g$ is the acceleration due to gravity, and $\nu$ is the kinematic viscosity of the fluid . We also use another parameter, reduced Rayleigh number $r=R/R_{c}$, where $R_{c}$ is the critical Rayleigh number.

The top and bottom boundaries are considered to be stress free and
perfectly conducting: 
\begin{eqnarray}
u_3 = \partial_{z}u_1 =\partial_{z}u_2 = \theta = 0, ~~~~ \mbox{at}~~  z = 0, 1.  \label{eq:bc} \end{eqnarray}
 We assume periodic boundary conditions along the horizontal direction ($x$ and $y$-direction). The boundary conditions chosen here are ideal. However they allow us to choose Fourier and $\sin$ or $\cos$ basis functions (SCFT) in our DNS. 
 
The amount of heat transported in the convection process is measured by the Nusselt number ($Nu$), which is defined as the ratio of total heat flux to the conductive heat flux. Using the nondimensionalization defined earlier, it can be shown
\begin{equation}
Nu = 1 + \langle u_{3}\theta\rangle, \label{nu}
\end{equation}
where, $\langle \cdot \rangle$ stands for spatial averaging. Please refer to Thual~\cite{Thual:JFM1992}, for a detailed derivation of the expression for Nusselt number.
 
Thual~\cite{Thual:JFM1992} numerically solved the Eqs.~\ref{eq:NS}-\ref{eq:continuity} under the above boundary conditions (Eq.~\ref{eq:bc})  using a pseudo-spectral code. His simulations were performed in a two-dimensional (2D) box (aspect ratio $\Gamma = 2\sqrt{2}$) for $r = 1.1$ to 70 and $P=6.8$.   To verify TARANG we compare Nusselt numbers obtained in our simulations with those of Thual's. The above set of Eqs.~(\ref{eq:NS}-\ref{eq:continuity})  are solved numerically using TARANG under the above boundary conditions (Eq.~\ref{eq:bc}). We use same geometry ({\em i.e.}\ 2D box with aspect ratio $2\sqrt{2}$) as used by Thual.  We use Fourier basis functions for
representation  along the $x$, and $\sin$ or $\cos$ functions for representation  along  the $z$ direction (SCFT basis). The validation results are shown in Table~\ref{rbc_table}. $Nu$ values obtained with TARANG are in very good agreement with those of Thual's until oscillation sets in the system.

\begin{table}[htbp]
\caption{Validation of TARANG against Thual's~\cite{Thual:JFM1992} 2D
simulations. We compare Nusselt numbers ($Nu$) obtained in our
simulations with grid resolution $64^{2}$ against Thual's  $16^{2}$
(THU1), $32^{2}$ (THU2), and $64^{2}$ (THU3) simulations. All $Nu$
values tabulated below are for $P=6.8$.}{

\begin{tabular}{ c c c c c }
\toprule
$r$ & THU1 & THU2 & THU3 & TARANG  \\[6pt]
\hline\\[-2pt]
2 & 2.142 & -- & -- & 2.142 \\
3 & 2.678 & -- & -- & 2.678 \\
4 & 3.040 & 3.040 & -- & 3.040 \\
6 & 3.553 & 3.553 & -- & 3.553 \\
10 & 4.247 & 4.244 & -- & 4.243 \\
20 & 5.363 & 5.333 & 5.333 & 5.333 \\
30 & 6.173 & 6.105 & 6.105 & 6.105 \\
40 & 6.848 & 6.742 & 6.740 & 6.740 \\
50 & 7.441 & 7.298 & 7.295 & 7.295 \\
70 & -- & oscil. & oscil. & 8.267 \\
\botrule
\end{tabular}}
\label{rbc_table}
\end{table}

\section{Parallelization}
TARANG has been organized in a modular manner, so parallelization of the code was quite straight forward.  Another major advantage was availability of parallel FFTW.  We essentially adopt FFTW's strategy for dividing the arrays etc.   If $p$ is the number of available processors, we divide each of the arrays into $p$ segments.  For example, a complex array $A(N_1, N_2, N_3/2+1)$ is split into $A(N_1/p, N_2, N_3/2+1)$ segments, each of which is handled by a processor.  The other major parallel tasks needed is the multiplication in real space, which is handled by individual processors.  Input/output is presently handled by the master node that collects/distributes data from the processor nodes.  We are planning to implement parallel input/output using HDF5 functions.

\section{Porting to GPU}
Graphics Processing Units (GPUs) are getting popular in high performance computing due to their larger number of cores. We have ported the FFT part of TARANG to GPUs, and have observed a reasonable speedup.  We are in the process of porting the entire code to multiple GPU platform.

\section{Future Plans}
We have successfully performed simulations for fluid, MHD, and convective flows for grids up to $1024^3$ on several platforms including PARAM YUVA (Centre for Advanced Computing, Pune), EKA (Computational Research Laboratory, Pune), HPC and CHAOS (both at IITK Kanpur).  Attempts are being made to run turbulence simulations on higher grids.

We have solved channel flow using Chebyshev and Fourier basis functions.  We will be porting the full implementation of the above basis function to be able to solve RBC and MHD flows under no-slip or mixed boundary conditions at the walls.  The other planned modules are flows for the cylindrical and spherical geometry.  We are also attempting to test and operationalize the magnetoconvection module.

\section{Conclusions}
TARANG exploits the object-oriented programming features of C++ to build flow solvers for incompressible fluid, MHD, and convection.  We adopt a modular approach where general purpose functions are assembled to create solvers for different situations.   This approach is proving to be very useful for constructing a large scale parallel softwares for fluid flows.

%\bibliographystyle{apsrev}
%\bibliography{/Users/mkv/res/bib/turbulence.bib}

\end{document}